%% file: paper-arxiv.tex
\documentclass[journal, onecolumn, 11 pt]{IEEEtran}

\newlength{\tocsep}
\usepackage{epsfig,float,multirow,footmisc,rotating}
\usepackage{subfigure}
\usepackage{graphics}
\usepackage{graphicx}
\usepackage{color}
\usepackage{subfig} 

\usepackage[utf8]{inputenc}
\usepackage{verbatim}
\usepackage{graphicx}
\usepackage{makeidx}
\usepackage{url}
\usepackage{array}
\DeclareGraphicsExtensions{.jpg, .png, .pdf, .tiff}

\usepackage{wrapfig}
\usepackage{sidecap}
\usepackage{float}
\usepackage[labelfont=bf]{caption}
\usepackage{multicol}

\usepackage{sidecap}
\usepackage{multimedia}
\usepackage[]{units}
\usepackage{amsmath,amsfonts,amssymb}
\usepackage[normalem]{ulem}
\usepackage{environ}
\usepackage{afterpage}

\definecolor{darkgreen}{RGB}{0,127,0}

\definecolor{darkred}{RGB}{127,0,0}

\definecolor{darkgrey}{RGB}{127,127,127}

\newcommand{\germanica}{\textit{B. germanica}}
\newcommand{\americana}{\textit{P. americana}}

\author{
Leo Cazenille$^{1,2}$,
Nicolas Bredeche$^{2}$,
Jos\'{e} Halloy$^{1}$
\\
{1}  Univ Paris Diderot, Sorbonne Paris Cit\'e, LIED, UMR 8236, 75013, Paris, France
\\
{2}  Sorbonne Universit\'es, UPMC Univ Paris 06, CNRS, ISIR, F-75005 Paris, France
}

\title{Automated optimisation of multi-level models of collective behaviour in a mixed society of animals and robots}

\begin{document}

\flushbottom
\maketitle
\thispagestyle{empty}

\begin{abstract}
Animal and robotic collective behaviours can exhibit complex dynamics that require multi-level descriptions.
Here, we are interested in developing a multi-level modeling framework for the use of robots  in studies about animal collective decision-making. In this context, using robots can be useful for validating models \textit{in silico}, inducing calibrated repetitive stimuli to trigger animal responses or modulating and controlling animal collective behaviour.
However, designing appropriate biomimetic robotic behaviour faces a major challenge: how to go from the collective decision dynamics observed with animals to an actual algorithmic implementation in robots. In previous work, this was mainly done by hand, often by taking inspiration from human-designed models.
Typically, models of behaviour are either macroscopic (differential equations of the population dynamics) or microscopic (explicit spatio-temporal state of each individual).
Only microscopic models can easily be implemented as robot controllers.
Here, we address the problem of automating the design of lower level description models that can be implemented in robots and exhibit the same collective dynamics as a given higher level model.
We apply evolutionary algorithms to simultaneously optimise the parameters of models accounting for different levels  of description. This methodology is applied to an experimentally validated shelter-selection problem solved  by gregarious insects and robots. We successfully design and calibrate automatically both a microscopic and a hybrid model exhibiting the same dynamics as a macroscopic one. Our framework can be used for multi-level modeling of collective behaviour in animal or robot populations and bio-hybrid systems.

\end{abstract}


\begin{multicols}{2}

\input{introduction}

\input{models}

\input{results}

\input{discussion}

\section*{Acknowledgment}
This work has been funded by EU-ICT project 'ASSISIbf', no 601074.
Experiments presented in this paper were carried out using the Grid'5000 testbed, supported by a scientific interest group hosted by Inria and including CNRS, RENATER and several Universities as well as other organizations (see https://www.grid5000.fr).


\end{multicols}

\clearpage
\input{Supplementary}

\end{document}

%% file: introduction.tex
\section{Introduction}
Groups of animals are able to reach consensus collectively, when presented with mutually exclusive alternatives. Over the years, scientists have compiled a large collection of dynamics observed in collective decision-making systems, based on experimental observations.
These systems can be very complex, and it can be challenging to build models that appropriately describe the observed behaviours.

Animals societies are systems with a very large parameter space. They can be modeled in numerous ways, using information about individual physiology, individual behaviour, group behaviour and features of the environment~\cite{mondada2013general,mermoud2012thesis}.
The collective behaviour of animals can be viewed as a complex system, and exhibit dynamics at several levels of organisation (hierarchical organisation).
One of difficulty in the modeling process is to find the appropriate level of details to integrate.

These models can typically be categorised in two groups, describing two different levels of abstraction: macroscopic and microscopic (cf Fig.~\ref{fig:multilevelCR}). There is a large number of studies, mainly in physics, that investigate the methods and applications of both groups of models, and the relations between them. Macroscopic models describe the system at the population level~\cite{camazine2003self}. They provide analytical and mathematical formalism of the dynamics of the system, but they are difficult to use in a straight-forward way as robotic controllers. Microscopic models describe explicitly the state of each individual agents (\textit{e.g.} Agent-based models of flocking, like the Vicsek model~\cite{vicsek1995novel}). They accurately capture the behaviour of individuals, and their relation with the environment, but they do not describe the collective dynamics of the system.

Both kind of models offer a different interpretation of the observed behaviour, and are complementary. Additionally, only microscopic models can be used to replay the observed behaviour of animals in computer simulations~\cite{halloy2007social,cazenille2015multi}.

In this paper, we are interested in modeling the use of robots to study collective decision-making in animals. Robots are useful for a number of reasons~\cite{floreano2008bio,mitri2013using}: validating models \textit{in silico}~\cite{garnier2009self}, inducing stimuli to observe animal feedback~\cite{sempo2006integration,Gribovskiy2010,landgraf2010biomimetic,bonnet2012437}, modulating animal collective behaviour~\cite{halloy2007social}, etc. 

However, designing appropriate robotic behaviour faces one major challenge: how to go from the collective decision dynamics observed with animals to an actual algorithmic implementation in robots. In previous work, this was done by hand. For example, \cite{halloy2007social} addresses modulation of collective decision-making in cockroaches using robots. They used observation to build a macroscopic model by hand, and program the robot's behaviour by hand. Though very promising, this approach revealed that designing the robots behaviour is very challenging and would benefit from automation.

The scientific question we address in this paper is thus the following: \textbf{how to automate the design of a model, that can be implemented in robots, that exhibit the same collective dynamics as a given macroscopic model}.

We propose a methodology to automate the design of a microscopic model, by using information both at the macroscopic level (a pre-existing macroscopic model) and at the microscopic level (pre-established knowledge of the individual behaviour of the animals). The objective of our methodology is to exhibit the same collective dynamics with this new microscopic model as predicted by the macroscopic model, improved with the ability to accurately simulate the microscopic interaction between individuals.

In the following, we apply our methodology to the collective decision-making problem described in~\cite{ame2006collegial,halloy2007social}, where a group of cockroaches must reach a consensus on choosing a preferred resting site (a \textit{shelter}). These papers introduced a general (macroscopic) Ordinary Differential Equations (ODE) model of the shelter-selection dynamics, that is experimentally validated. Starting from the ODE model,  we make use of experimental data of individual cockroach behaviour from~\cite{jeanson2003model,garnier2009self} as \textit{a-priori} microscopic information. From these two sources of information, we show how our method can be used to calibrate a target model.
We consider two (target) models: a Finite State Machine (FSM) agent-based microscopic model~\cite{cazenille2015multi}, and an agent-based Hybrid model, combining macroscopic and microscopic information, that was already used with manually-defined parameters in~\cite{halloy2007social}.
Validation of these models is achieved by comparing their shelter-selection dynamics to those exhibited by the ODE model.
The parameters of the FSM and Hybrid models are automatically calibrated so that they exhibit the same collective dynamics as ODE model, by using evolutionary algorithms.

%% file: models.tex
\section{Models for collective decision-making by cockroaches} \label{sec:models}
\begin{figure}[H]
	\centering
	\includegraphics[width=8.3cm]{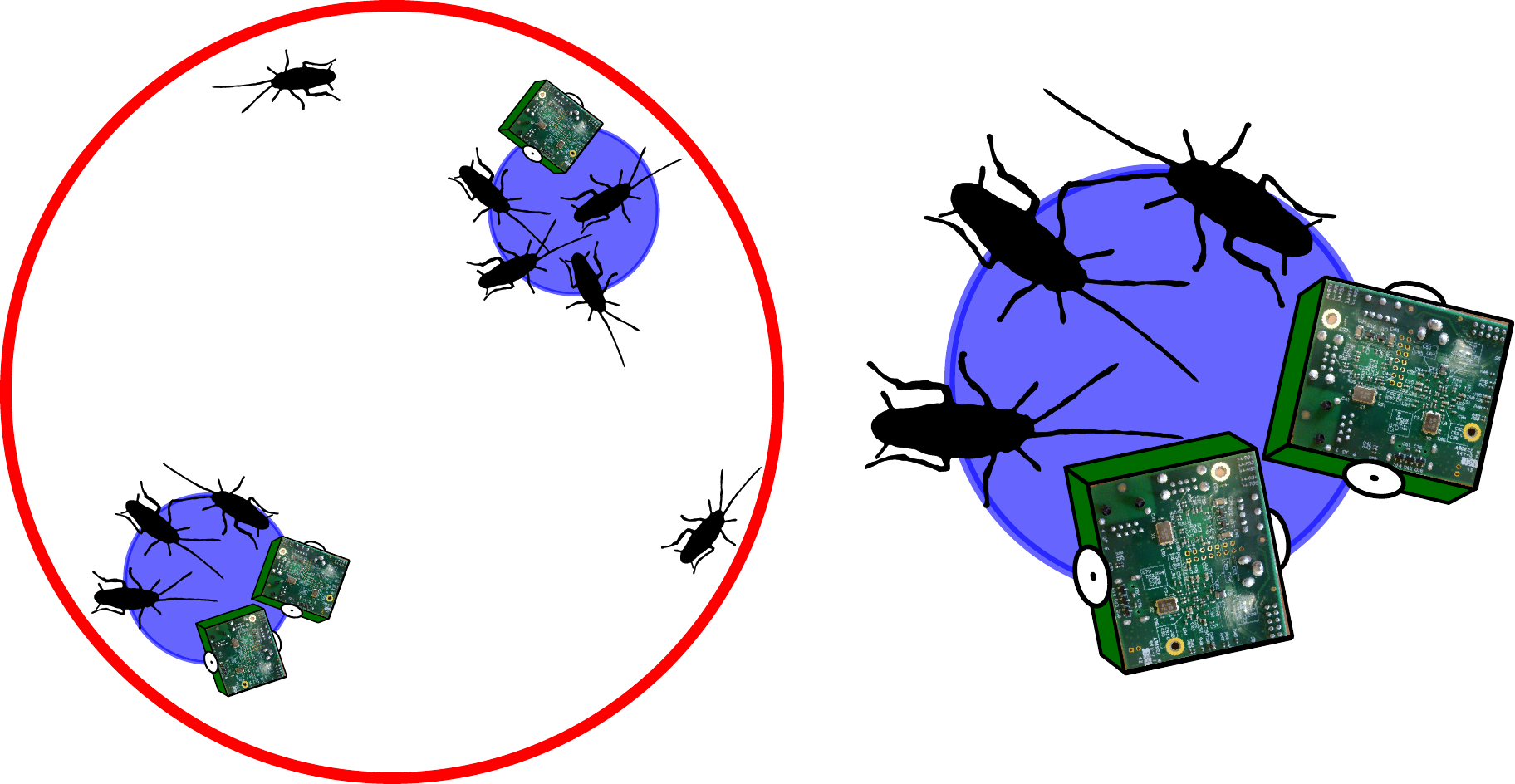}
	\caption{\textbf{Experimental setup used in~\cite{halloy2007social}}. It includes two identical shelters and both cockroaches (\americana, approximate size: $\sim 4cm$, surface: $600mm^2$, or \germanica, size: $\sim 0.25cm$, surface: $3mm^2$) and robots (surface: $1230mm^2$ in \americana\  setups, $6.15mm^2$ in \germanica\  setups) in a circular arena (diameter: $1m$ for \americana, $14cm$ for \germanica). The setup is symmetric.}
\label{fig:ArenaCRExpe}
\end{figure}

We simulate the experimental setup from~\cite{ame2006collegial,halloy2007social} (cf Fig.~\ref{fig:ArenaCRExpe}): cockroaches are put into a circular arena with two identical shelters (resting sites). In this setup, it has been shown that cockroaches tend to aggregate under the shelters. Two species of cockroaches are considered: \americana\  and \germanica. This setup is adapted to study collective decision-making, because it imply a trade-off between cooperation (aggregation of the individuals) and competition for resources with limited carrying capacity (the shelters). 

We consider three models of cockroaches collective behaviour in a shelter selection problem (cf Fig.~\ref{fig:tableModels}): a macroscopic ODE model, a microscopic FSM model, and an hybrid model combining macroscopic and microscopic levels of abstraction. All three models can handle time discrete data. The ODE model does not include extended spatial information of the individuals. The FSM and Hybrid models include explicit spatial information. As they also possess a microscopic component, they could be implemented as robotic controllers.
A classification of models according to their level of abstraction can be found in~\cite{mermoud2012thesis}.

\begin{figure}[H]
\centering
\includegraphics[width=8.0cm]{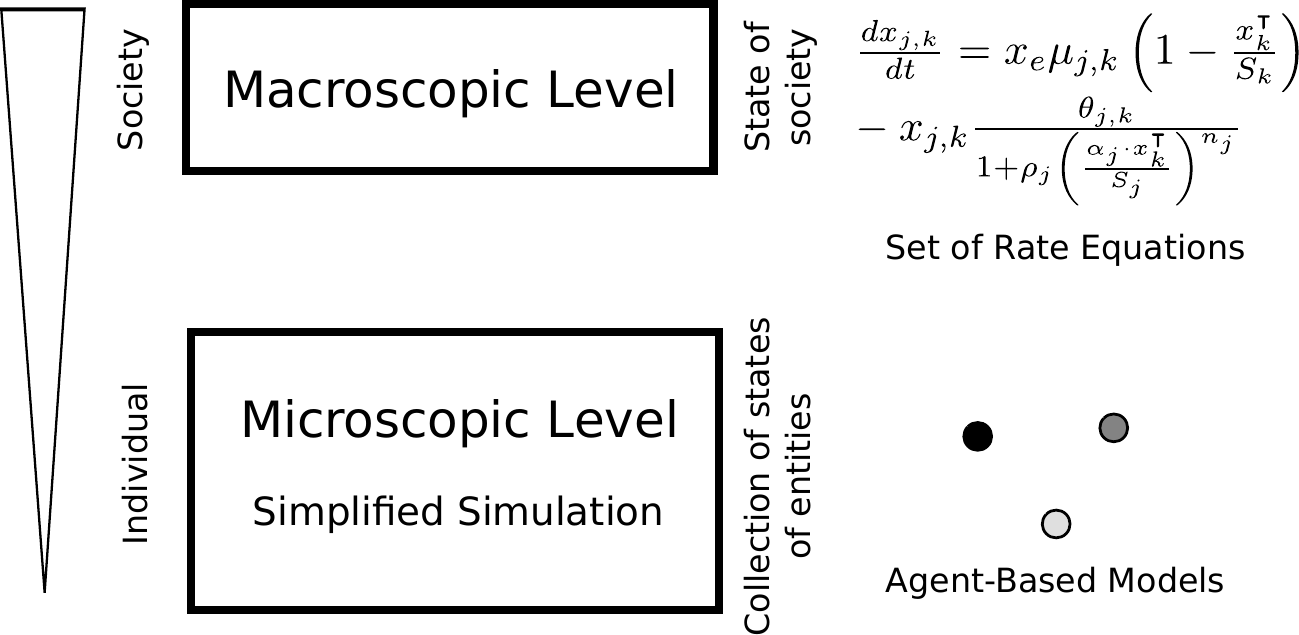}
\caption{\textbf{Difference between macroscopic and microscopic models} Models of stochastic and dynamical systems can tread upon different levels of abstraction. These models can typically be separated in two groups: macroscopic and microscopic. Macroscopic models describe the global state of the system (i.e. the state of the society). Macroscopic models usually take the form of sets of equations (\textit{e.g.} ODE) by doing a mean field hypothesis. Microscopic models describe the state of each individual agent. They usually are Agent-Based models (ABM). Agent motion strategy is often represented as Finite State Machine (FSM). Macroscopic and microscopic models complement each other, as they describe dynamics at different levels. Microscopic models can include information about the spatiality of the agents, which enable them to be used to replay the modeled behaviour in simulations.}
\label{fig:multilevelCR}
\end{figure}

\end{multicols}
\clearpage

\begin{figure}[H]
\small
\centering
\begin{tabular}{ c  c  c  c }
\hline
Name & Experimentally validated & Wall-Following behaviour & Constant Speed \\
\hline
ODE & yes~\cite{ame2006collegial, halloy2007social} & no & yes \\
FSM & yes~\cite{jeanson2003model, garnier2009self} & yes & no \\
Hybrid & partially~\cite{halloy2007social} & no & yes \\
\hline
\end{tabular}
\caption{\textbf{Comparative table of studied models}. The ODE model is a Mean field description of the problem. The FSM model is an agent-based model using a Finite State Machine representation. The hybrid model combine macroscopic information (inspired from the ODE model), and microscopic information (with an approach similar to the FSM model).}
\label{fig:tableModels}
\end{figure}
\begin{multicols}{2}

\subsection{Mean field description: Ordinary Differential Equation model} \label{sec:ode}
Halloy \textit{et al.}~\cite{halloy2007social} describe a mathematical model of the collective dynamics of mixed groups of cockroaches and robots in a shelter-selection problem (from~\cite{ame2006collegial}). In this model, animals and robots influence equivalently the collective decision-making process, and they exhibit homogeneous behaviour. This model handles two populations (animals and robots) in setups with two shelters.
The following set of ODE represents the evolution of the number of individuals in each shelter (and outside), in a setup with two shelters:

\begin{equation} \label{eq:animalsEq}
	\frac{d x_{i}}{d t} =
		x_e \mu_i \left(1 - \frac{x_i + \omega r_i}{S_i} \right) -
		x_i \frac{\theta_i}{1 + \rho \frac{x_i + \beta r_i}{S_i}^n }
\end{equation}
\begin{equation} \label{eq:robotsEq}
	\frac{d r_{i}}{d t} =
		r_e \mu_{ri} \left(1 - \frac{x_i + \omega r_i}{S_i} \right) -
		r_i \frac{\theta_{ri}}{1 + \rho_r \frac{\gamma x_i + \delta r_i}{S_i}^{n_r} }
\end{equation}

\begin{equation} \label{eq:totEq}
	C = x_e + x_1 + x_2, \quad
	R = r_e + r_1 + r_2, \quad
	M = R + C
\end{equation}

Table~\ref{fig:parametersModels} describes the parameters of this ODE model.

The influence of animals on animals ($\alpha$) is equal to $1.0$, and is not considered in~\cite{halloy2007social}: the assumption is made that this parameter is imposed by biology, and can't be changed in experiments.
Because of crowding effects, the probability of an individual joining a shelter decreases with the level of occupancy of this shelter.
We assume the two shelters to have the same carrying capacity: $S = S_1 = S_2$.
We define a measure: $\sigma = S/C$, corresponding to the sites carrying capacity as a multiple of the total population.

\begin{figure}[H]
\centering
	\includegraphics[width=15.10cm]{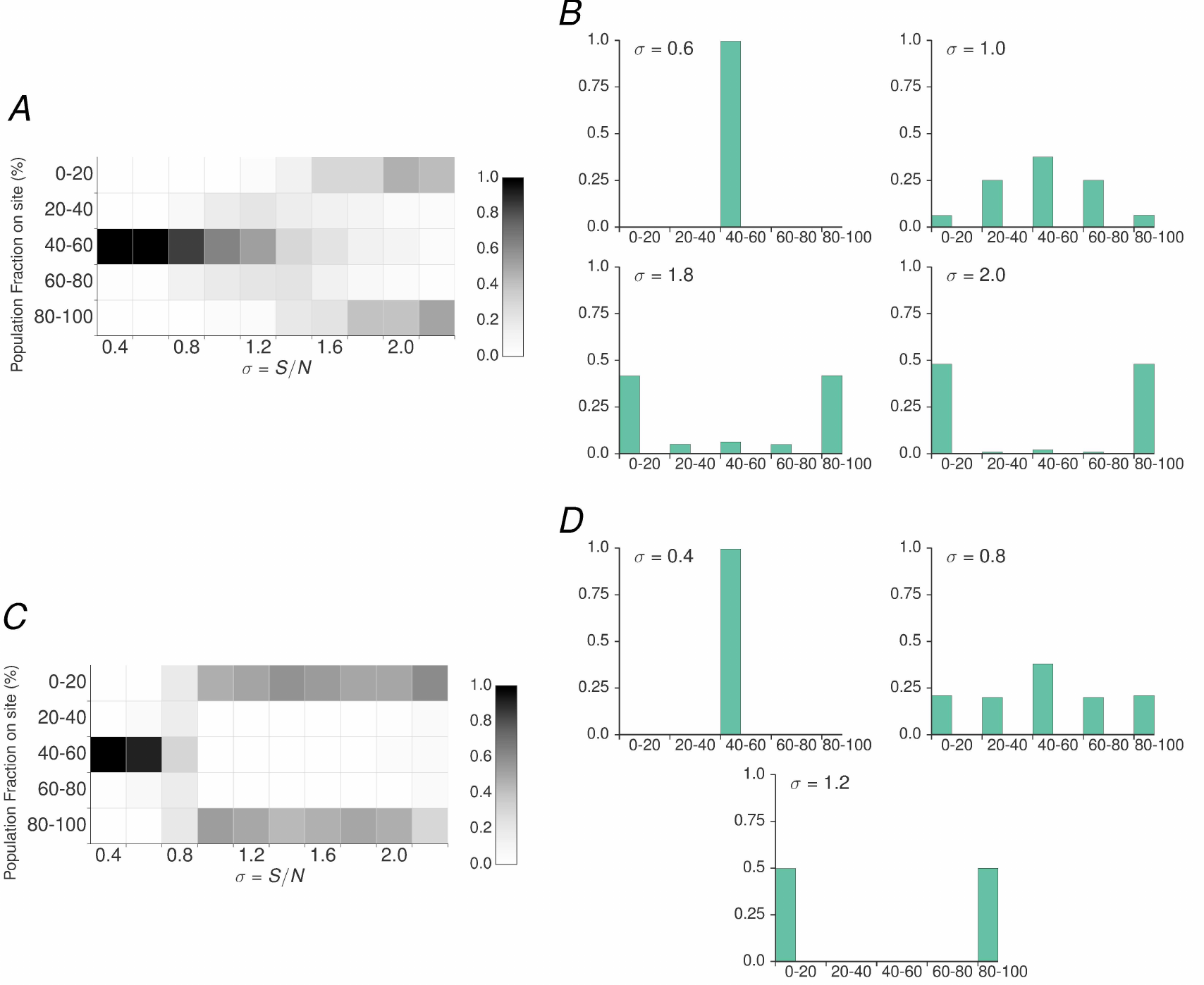}
	\caption{\textbf{Bifurcation diagrams and distributions of $M=50$ \germanica\  (Panels A and B) or \americana\  (Panels C and D) cockroaches in the first shelter, as function of $\sigma$~\cite{ame2006collegial}}.
	The bifurcation diagrams are represented as bi-dimensional histograms of the results of using $1000$ solutions by parameter sets.
	The colour of each bin of the histograms corresponds to the occurrence of experiments.
	The diagrams are symmetric for all tested values of $\sigma$, so only one shelter is represented.
	The bifurcation point is close to $\sigma = 0.8$ for \americana, and $\sigma = 1.0$ for \germanica.
	Before the bifurcation point ($0.4 \leq \sigma < 0.8$ for \americana, $0.4 \leq \sigma < 1.0$ for \germanica), only one configuration exists, corresponding of an equipartition of the individuals ($x_1/C = x_2/C = 1/2, x_e=0$).
	After the bifurcation point ($\sigma > 0.8$ for \americana, $\sigma > 1.0$ for \americana), two stable configurations exist, corresponding to all individuals in one of the shelters (either $x_1 \approx 0, x_2 \approx 1, x_e \approx 0$ or $x_1 \approx 1, x_2 \approx 0, x_e \approx 0$).
}
\label{fig:bifurcationsCR}
\end{figure}

When no robots are present ($R = 0$) and only animals are considered, two different dynamics are observed. The bifurcation point is close to $\sigma = 0.8$ for \americana, and $\sigma = 1.0$ for \germanica.
Before the bifurcation point ($0.4 \leq \sigma < 0.8$ for \americana, $0.4 \leq \sigma < 1.0$ for \germanica), only one configuration exists, corresponding of an equipartition of the individuals ($x_1/C = x_2/C = 1/2, x_e=0$).
After the bifurcation point ($\sigma > 0.8$ for \americana, $\sigma > 1.0$ for \americana), two stable configurations exist, corresponding to all individuals in one of the shelters (either $x_1 \approx 0, x_2 \approx 1, x_e \approx 0$ or $x_1 \approx 1, x_2 \approx 0, x_e \approx 0$)~\cite{ame2006collegial}.
Only results with population of $50$ cockroaches are represented in Fig.~\ref{fig:bifurcationsCR}, but similar dynamics are observed with different population sizes.

\end{multicols}
\clearpage

\begin{figure}[htp!]
	\footnotesize
	\centering
	\begin{tabular}{@{} p{0.2cm} c c c c c @{}}
		\cline{2-6}
		& Parameter & Value for \americana & Value for \germanica & Optimised & Description \\
		\cline{2-6}
		& $P$	& \multicolumn{2}{c}{$2$} & & Number of sites\\
		& $S_i$ & & & & Carrying capacity of shelter $i$\\
		& $C$   & \multicolumn{2}{c}{$50$} & & Number of agents\\
		& $x_i$ & & & & Number of agents in shelter $i$\\
		& $x_e$ & & & & Number of agents outside the shelters\\
		\cline{2-6}

		\multicolumn{1}{c}{\parbox[c][0.0cm][t]{2mm}{\multirow{3}{*}{\rotatebox[origin=c]{90}{ODE}}}}
		& $\mu_i$ & $0.0027 s^{-1}$ & $0.001 s^{-1}$ & & Maximal kinetic constant of entering a shelter\\
		\multicolumn{1}{c}{} & $\theta_i$ & $0.44 s^{-1}$ & $0.01 s^{-1}$ & & Maximal rate of leaving a shelter\\
		\multicolumn{1}{c}{} & $\rho$, $n$ & $4193$, $2.0$ & $1667$, $2.0$ & & Influence of conspecifics\\
		\cline{2-6}

		\multicolumn{1}{c}{\parbox[c][0.0cm][c]{2mm}{\multirow{9}{*}{\rotatebox[origin=c]{90}{FSM}}}}
			& $l_c$			& \multicolumn{2}{c}{$\left[1.0, 500.0\right] cm$} & yes & Mean size of path\\
		\multicolumn{1}{c}{} & $a_c$			& \multicolumn{2}{c}{$\left[-\pi, \pi\right]$} & yes & Geometric mean for angle departure\\
		\multicolumn{1}{c}{} & $\tau_{c,exit}$	& \multicolumn{2}{c}{$\left]0.0, 10.0\right[ s$} & yes & Mean time an agent follow a wall\\
		\multicolumn{1}{c}{} & $v_{c,c}$		& \multicolumn{2}{c}{$\left]0.0, 3.0\right[ cm.s^1$} & yes & Mean speed in central zone\\
		\multicolumn{1}{c}{} & $v_{c,p}$		& \multicolumn{2}{c}{$\left]0.0, 3.0\right[ cm.s^1$} & yes & Mean speed in peripheral zone\\
		\multicolumn{1}{c}{} & $s_{c,i,n}$		& \multicolumn{2}{c}{$\left[0.0, 1.0\right]$} & yes & Probability of stop in shelter $i$ with $n$ neighbors\\
		\multicolumn{1}{c}{} & $\tau_{c,i,n}$	& \multicolumn{2}{c}{$\left]0.0, 1000.0\right[ s$} & yes & Mean stop duration in shelter $i$ with $n$ neighbors\\
		\multicolumn{1}{c}{} & $d$	& \multicolumn{2}{c}{$\left]0.8, 1.0\right[ m$} & yes & Diameter of the central zone\\
		\cline{2-6}

		\multicolumn{1}{c}{\parbox[c][0.0cm][c]{2mm}{\multirow{6}{*}{\rotatebox[origin=c]{90}{Hybrid}}}}
			& $\theta_i$ & \multicolumn{2}{c}{$\left]0.0, 0.50 \right] s^{-1}$} & yes & Maximal rate of leaving a shelter\\
		\multicolumn{1}{c}{} & $\rho$, $n$ & \multicolumn{2}{c}{$\left[500, 5000 \right]$, $2.0$} & yes & Influence of conspecifics\\
		\multicolumn{1}{c}{} & $l$		   & \multicolumn{2}{c}{$\left[1.0, 500.0\right] cm$} & yes & Mean size of path\\
		\multicolumn{1}{c}{} & $a$		   & \multicolumn{2}{c}{$\left[-\pi, \pi\right]$} & yes & Geometric mean for angle departure\\
		\multicolumn{1}{c}{} & $v$		   & \multicolumn{2}{c}{$\left]0.0, 3.0\right[ cm.s^{-1}$} & yes & Constant speed of agents\\
		\cline{2-6}

	\end{tabular}

	\caption{\textbf{Parameters list of the ODE, FSM and Hybrid models. Cockroaches (\americana\  and \germanica) parameter values for the ODE model are from~\cite{halloy2007social} and~\cite{ame2006collegial}.} We only consider the case where $M = 50$. In setups with two shelters, the ODE, FSM and Hybrid models have respectively $18$, $45$ and $20$ parameters. Parameter values used for the FSM and Hybrid models are obtained in a calibration process described in Sec.~\ref{sec:results} and Supplementary~\ref{sec:supplementaryCalibration}. The influence of animals on animals ($\alpha$) is equal to $1.0$, and is not considered in~\cite{halloy2007social}: the assumption is made that this parameter is imposed by biology, and cannot be changed in experiments. All parameters of the FSM and Hybrid models are optimised, using the method described in Supplementary~\ref{sec:supplementaryCalibration}, to calibrate these models so that they exhibit the same collective dynamics as those described in the reference ODE model.}
\label{fig:parametersModels}
\end{figure}


\begin{figure}[htp!]
	\centering
	\includegraphics[width=13cm]{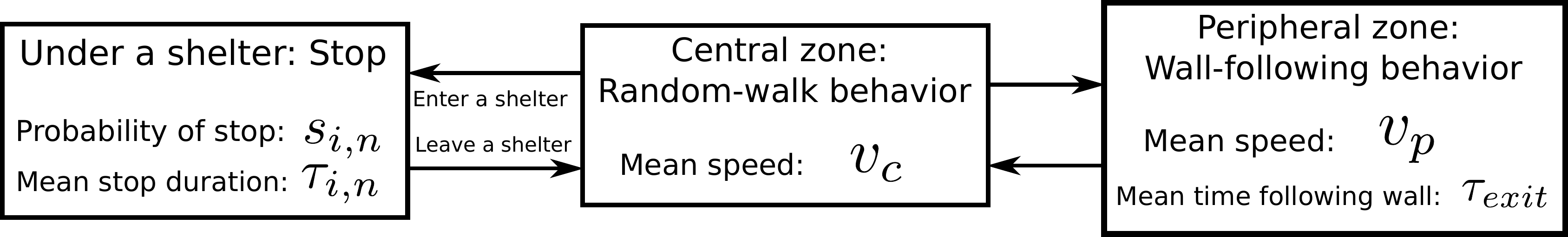}

	\caption{\textbf{Finite State Machine model of cockroach individual behaviour.} The arena contains two zones: the peripheral zone (agents display a wall-following behaviour), and the central zone (agents display a random-walk behaviour). Shelters are in the central zone. When an agent enters a shelter, it has a probability of stopping for a random duration before exiting the shelter. The probability of stopping under shelter depends on the number of neighbors present in the shelter, and can be different for each shelter. Only $10$ neighbors are considered in our experiments. In setups with two shelters, this model has $45$ parameters per population.}
	\label{fig:microModel}
\end{figure}

\begin{multicols}{2}

\subsection{Finite State Machine model}
We consider the Finite State Machine described in~\cite{cazenille2015multi}, as agent-based model of cockroaches and robots behaviour. This model is inspired from the agent-based aggregation models in~\cite{jeanson2003model,garnier2009self}, that describe the collective behaviour of cockroaches in a similar setup.

Cockroaches have a tendency to follow walls when they are already close to a wall of the arena. The model defines two zones in the arena: the ring area that borders the walls of the arena is called the \textit{peripheral zone}, while the rest of the arena corresponds to the \textit{central zone}. In the \textit{peripheral zone}, agents follow a wall-following behaviour. In the central zone, agents follow a random-walk behaviour, with trajectories composed of a recurring alternation of straight lines and rotations.  Shelters are all in the central zone. We do not model the actual trajectories of cockroaches.

\newpage
When agents enter a shelter, they have a probability of stopping for a random timespan before moving away from the shelter. Similarly to~\cite{garnier2009self}, this probability depends on the number of agents present under the shelter, as cockroaches are gregarious during their resting period. However, in this model (as opposed to~\cite{jeanson2003model,garnier2009self}), the probability of stopping when under a shelter is not the same for both shelters: this model is more general, and can be useful when describing more complex behaviours with asymmetric decision-making dynamics~\cite{cazenille2015multi}. 

Figure~\ref{fig:microModel} represents the Finite State Machine used in this model. The relevant model parameters are found in Fig.~\ref{fig:parametersModels}.

\end{multicols}

\clearpage
\begin{multicols}{2}

\subsection{Hybrid model}
We introduce a model of collective behaviour of cockroaches, using information at both macroscopic and microscopic levels of abstraction. We call this multi-level model a "Hybrid model".

\begin{figure}[H]
	\centering
	\includegraphics[width=8.3cm]{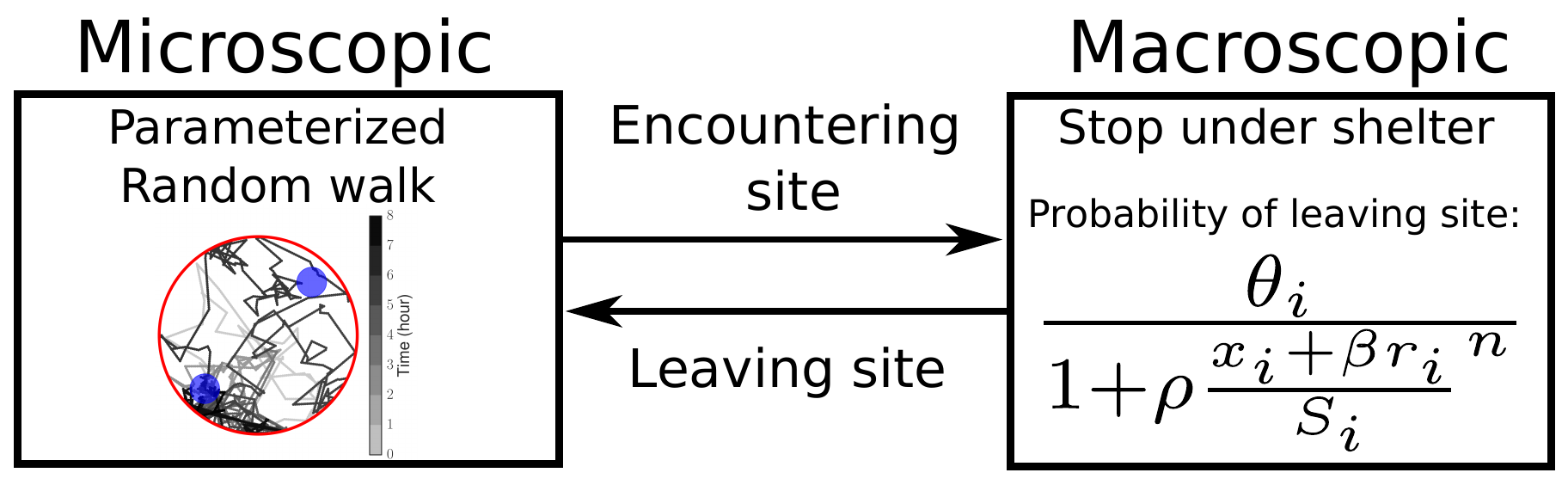}
	\caption{\textbf{Finite State Machine representation of the hybrid behavioural model of one agent.} Table~\ref{fig:parametersModels} references the parameters of the model. The model describes 2 kind of behaviour: when the agents are not under a site, they will exhibit a random-walk behaviour, following a recurring alteration of straight lines and rotations. This behaviour is microscopic, because agents only use local information to determine their course of action. When agents encounter site, they will stop. At each subsequent time-step, the stopped agent has a probability of $\frac{\theta_i}{1 + \rho \frac{x_i + \beta r_i}{S_i}^n }$ (for cockroaches) or $\frac{\theta_{ri}}{1 + \rho_r \frac{\gamma x_i + \delta r_i}{S_i}^{n_r} }$ (for robots) of leaving site, and returning to the random-walk behaviour. This behaviour is macroscopic, as it require information about the whole group of agents. As the model combines microscopic and macroscopic parts, it is a multi-level (or hybrid) model.}
	\label{fig:hybridModel}
\end{figure}

Like the FSM model presented earlier, this model take into account the spatiality of the agents, and can be implemented in a robotic controller.
However, this model builds from the ODE model, so that the calibration process can make use of a relevant parameter space (described in Sec.~\ref{sec:results} and Supplementary~\ref{sec:supplementaryCalibration}), compared to the FSM model. As such, this hybrid model is a crossover between the macroscopic ODE model, which describes easily collective behaviour and site occupation, and the microscopic FSM model, which details the spatio-temporal behaviour of single agents.
Figure~\ref{fig:hybridModel} describes the hybrid model by using a Finite State Machine representation of the behaviour of a single agent.

When the agents are not under a shelter, they follow a random-walk behaviour (microscopic behaviour). When agents enter a shelter, they stop, and have a probability, at each subsequent time-step, of leaving the shelter. This probability, taken from Eq.~\ref{eq:animalsEq} and~\ref{eq:robotsEq}, is computed using macroscopic information, and is defined as such:
\begin{equation} \label{eq:hybrid}
	\frac{\theta_{i}}{1 + \rho \frac{\alpha x_i}{S_i}^{n} }
\end{equation}
To simplify the model, we did not include a wall-following behaviour, like the FSM model, and agents move with a constant speed.

\begin{figure}[H]
	\centering
	\includegraphics[height=3.6cm]{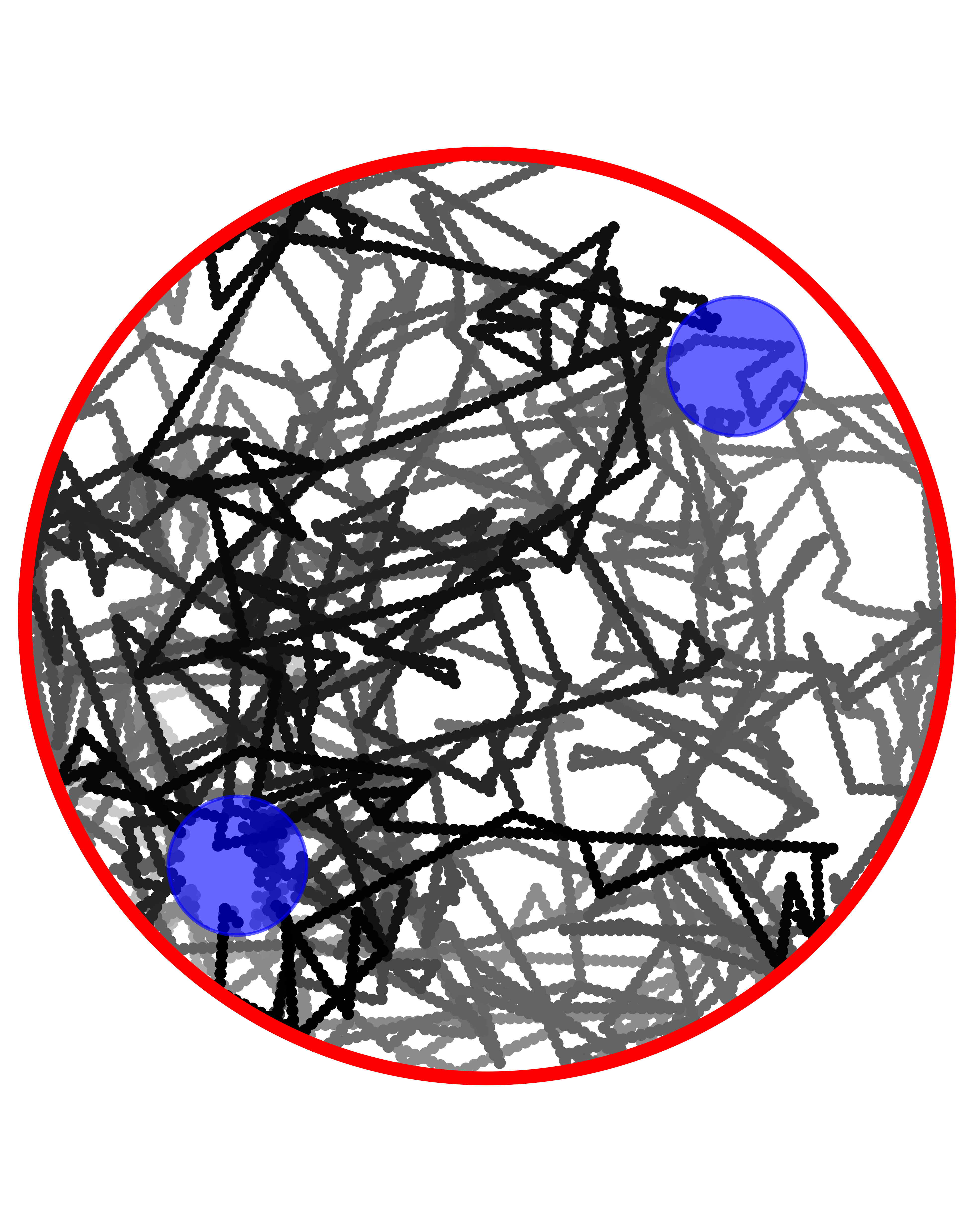}
	\includegraphics[height=3.6cm]{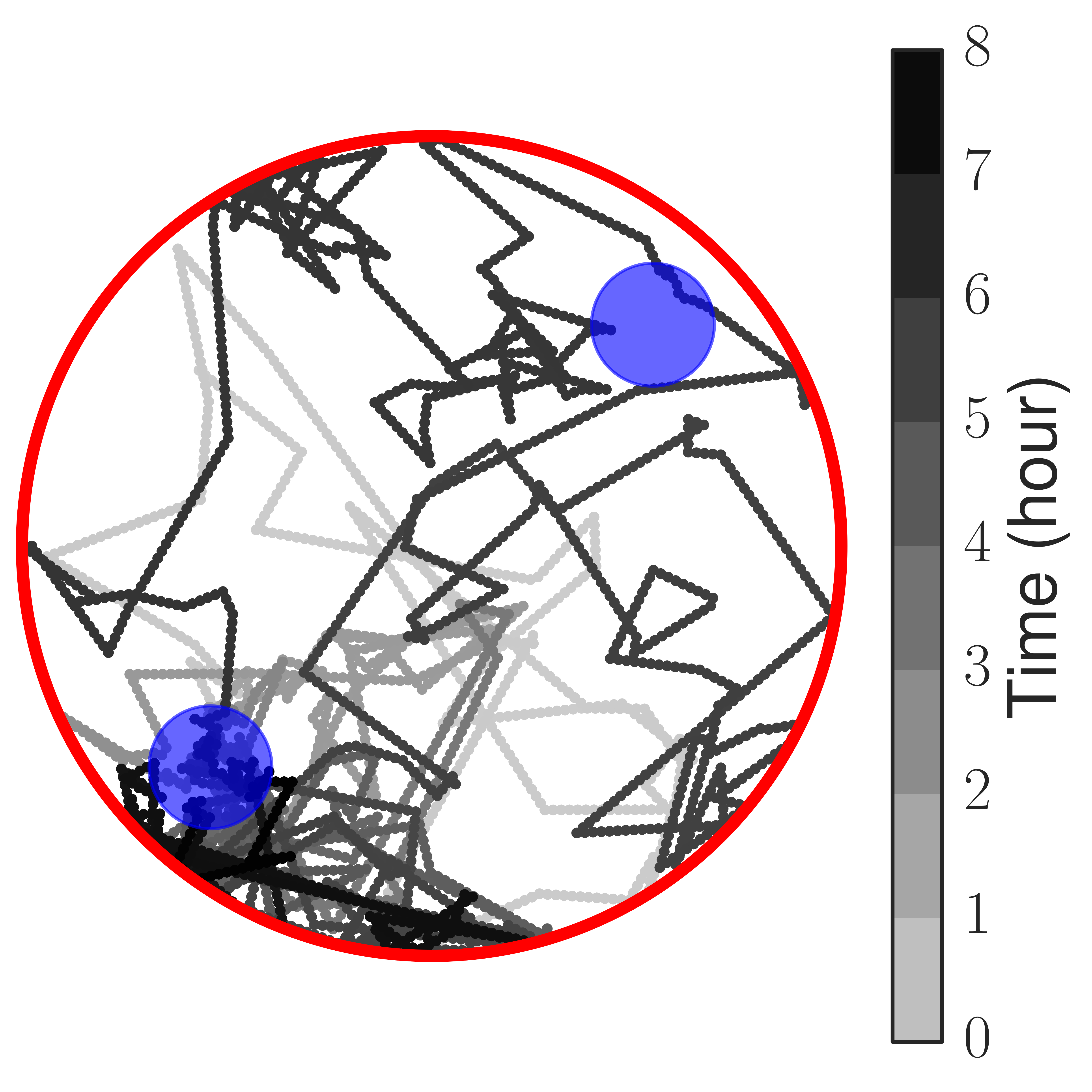}
	\caption{\textbf{Examples of the trajectory of a simulated cockroach, using the Hybrid model, in a population of $50$ cockroaches.} The arena is circular (diameter $1$ m) and contains two sites ($150$mm). Gray lines represents the (random-walk) trajectory of one agent. These trajectories are not meant to fit the normal trajectories of actual cockroaches: we designed our models to reproduce the observed random exploration. The brightness of the line reflects to simulation time. All experiments last $8$ hours.}
	\label{fig:traceHybrid}
\end{figure}

Figure~\ref{fig:traceHybrid} presents examples of the trajectories of single cockroaches in a simulation with population of $50$ cockroaches that follow the hybrid model.

%% file: results.tex
\section{Results} \label{sec:results}
\newcommand{\fitness}{\text{Fitness}}
\newcommand{\hellinger}{\text{Hellinger}}
\newcommand{\optimized}{\text{optimised}}
\newcommand{\reference}{\text{reference}}
\newcommand{\calibration}{\text{calibration}}
\newcommand{\Hist}{\text{Hist}}

Our goal is to find parameter sets of the FSM and Hybrid models describing robot behaviour that mimic the decision-making dynamics of the animals in a mixed-society. We consider two types of simulations, for both \americana\ and \germanica \   cockroach species. The first type describes a purely biological system, with only $50$ cockroaches (either \americana\ or \germanica) and no robots. It is used as the biological reference case. The second type is devoted to biohybrid groups made of $45$ cockroaches (either \americana\ or \germanica) and $5$ robots. The number of robots is kept small to reflect the settings used in a mixed-society experiment~\cite{halloy2007social} where a minority of robots is capable of controlling the whole mixed group behaviour.

\begin{figure}[H]
	\centering
		\includegraphics[width=12.00cm]{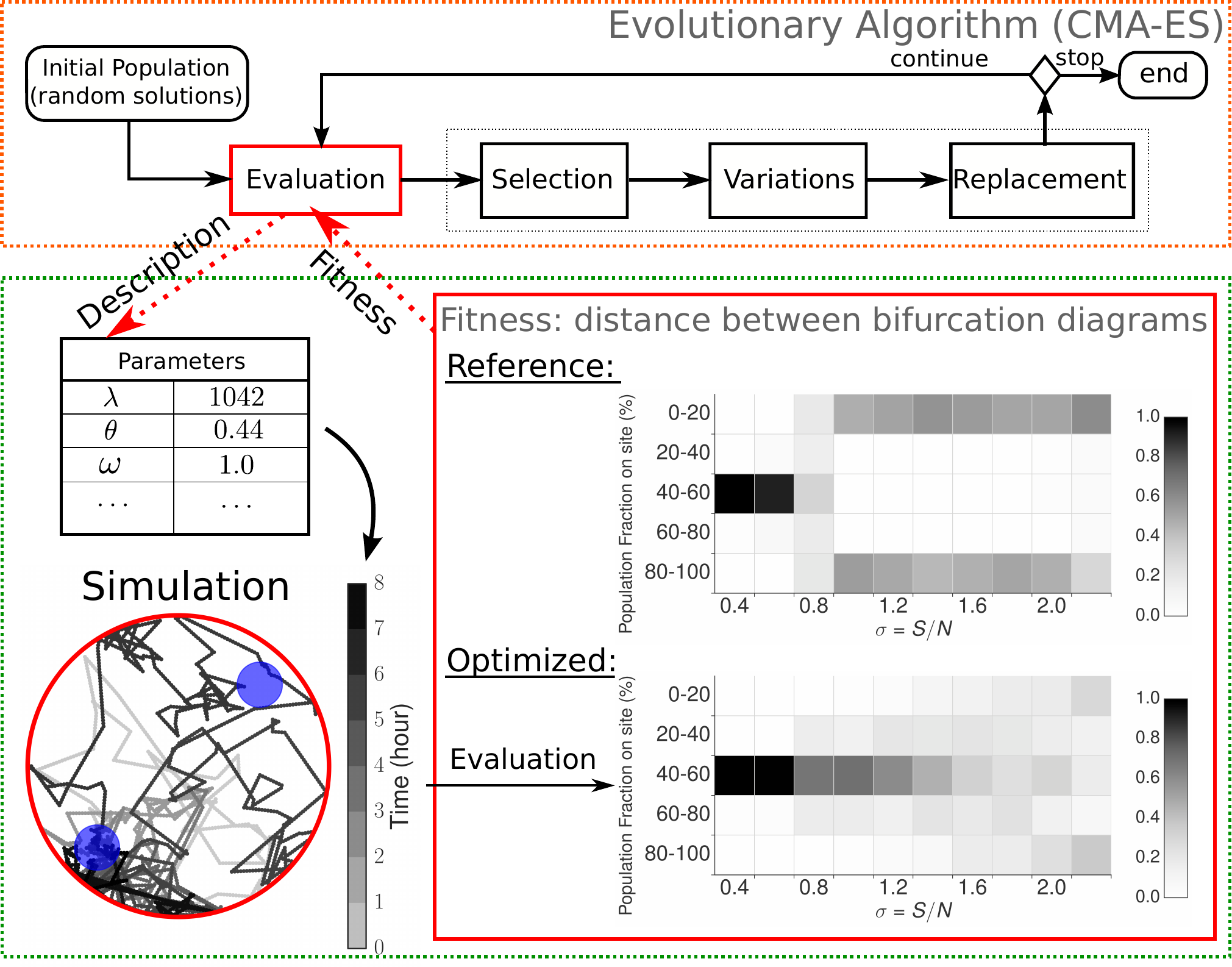}
	\caption{\textbf{Workflow of the Automated model calibration task by optimisation}. The optimised bifurcation diagram and the reference bifurcation diagram are both converted to one-dimensional histograms, by normalising the sum of all bin values to $1.0$. The optimiser will maximise the fitness, which is computed by the formula: $Fitness = 1.0 - D_{hellinger}(B_{optimized} / N_u, B_{reference} / N_u)$ where $N_u$ is the number of columns in the bifurcation diagrams ($10$) and $B_{optimized}$ and $B_{reference}$ are one-dimensional histograms version of the bifurcation diagrams. The term $N_u$ is used for normalisation. $D_{hellinger}(P, Q) = \sqrt{ 2 \sum_{i=1}^{d} (\sqrt{P_i} - \sqrt{P_i}  )^2 }$ is the Hellinger distance~\cite{deza2006dictionary}. This approach is described in Supplementary Information~\ref{sec:supplementaryCalibration}}
	\label{fig:optim-task0}
	\end{figure}

To use the FSM and Hybrid models describing animal or robot behaviour in simulation, we must calibrate them to exhibit the same decision-making dynamics as the ODE model.
As the ODE model is parameterised using experimental data, it allows the FSM and Hybrid models to describe as accurately as possible the (macroscopic) site-selection dynamics of the cockroaches. The calibration process is described in Fig.~\ref{fig:optim-task0}.

We optimise the parameter sets of the cockroaches individuals, for the FSM and Hybrid models: Fig.~\ref{fig:parametersModels} lists the parameters of these two models that are optimised.
Instances of the FSM and Hybrid models using these parameter sets are simulated for different values of $\sigma$. This allows to create bifurcation diagrams to be created for each optimised individuals, similar to Fig.~\ref{fig:bifurcationsCR}.

As there is only few \textit{a-priori} information about the parameter space, and as the parameter space has a relatively large dimensionality, we use the state-of-the-art CMA-ES evolutionary optimisation method~\cite{auger2005restart} to optimise the parameter sets of the FSM and Hybrid models. To evaluate the difference between two parameter sets, we use a distance metric between the two resulting bifurcation diagrams. This method is described in Supplementary Information~\ref{sec:supplementaryCalibration}.

\begin{figure}[htp]
\centering
\includegraphics[width=15.00cm]{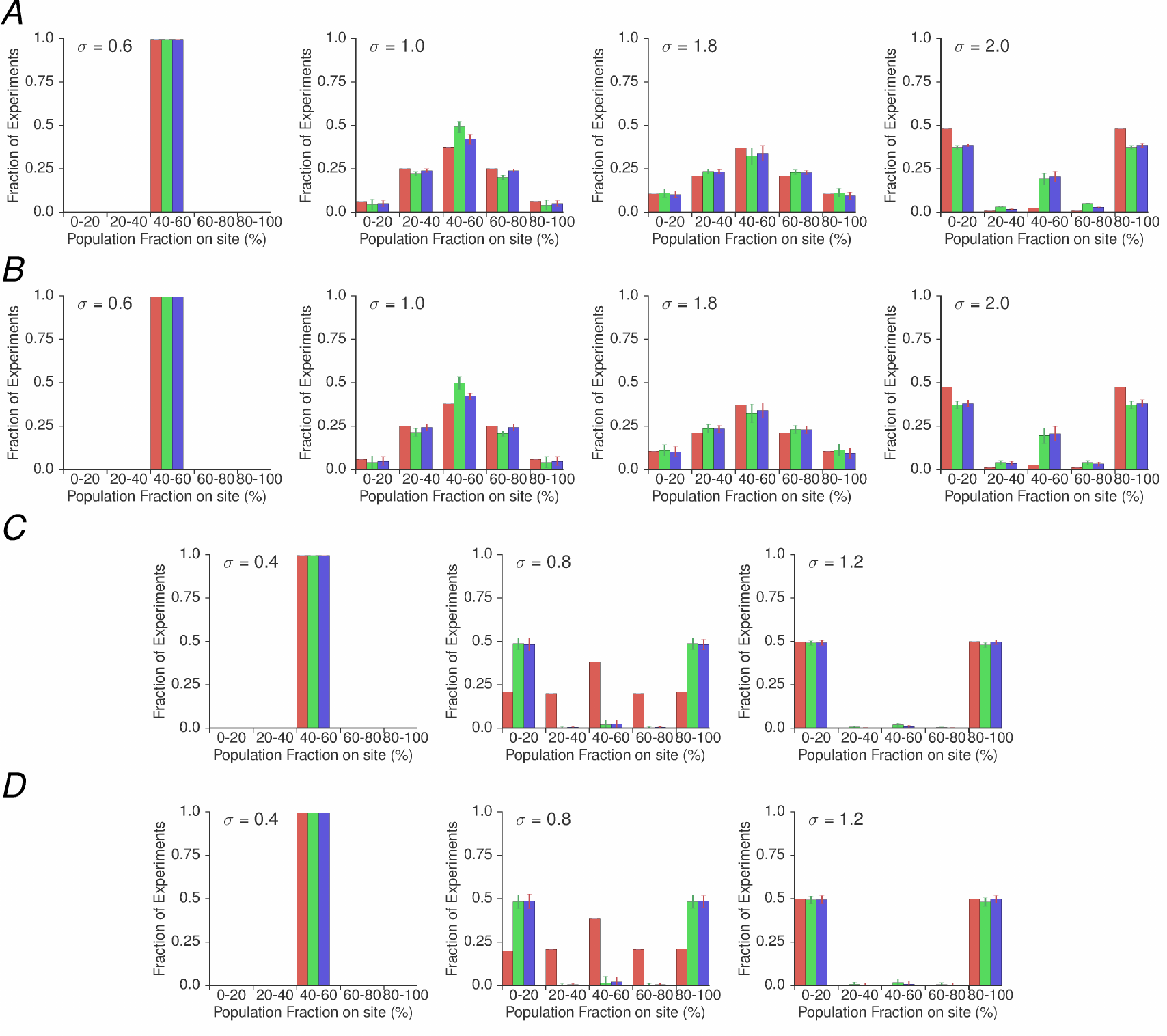}
\caption{\textbf{Distribution of $50$ agents in the first shelter for chosen values of $\sigma$, using three different models: ODE, FSM and Hybrid.} Panels A and C are obtained from simulations using $50$ cockroaches (Panel A: \americana\ , and Panel C: \germanica). Panels B and D are obtained from simulations using $45$ cockroaches and $5$ robots (Panel B: \americana\ , and Panel D: \germanica). ODE, FSM and Hybrid models results are respectively in red, green, and blue.
The bifurcation point is close to $\sigma = 0.8$ for \americana, and $\sigma = 1.0$ for \germanica. The parameter $\sigma$ values are chosen before the bifurcation point ($\sigma=0.4$ for \americana\ , $\sigma=0.6$ for \germanica), and just after the bifurcation point ($\sigma=1.2$ for \americana\ , $\sigma=1.8$ and $\sigma=2.0$ for \germanica). The best sets of optimised model parameters are used, after 100 runs of optimisation.
The diagram is symmetric for all tested values of $\sigma$, so only one shelter is represented.
Calibrated versions of the FSM and Hybrid models behave similarly to the ODE model: (1) before the bifurcation point ($0.4 \leq \sigma < 0.8$ for \americana, $0.4 \leq \sigma < 1.0$ for \germanica), only one configuration exists, corresponding of an equipartition of the individuals ($x_1/N = x_2/N = 1/2, x_e=0$); (2) after the bifurcation point ($\sigma > 0.8$ for \americana, $\sigma > 1.0$ for \americana), two stable configurations exist, corresponding to all individuals in one of the shelters (either $x_1 \approx 0, x_2 \approx 1, x_e \approx 0$ or $x_1 \approx 1, x_2 \approx 0, x_e \approx 0$).
}
\label{fig:calibrationRes}
\end{figure}

Figure~\ref{fig:calibrationRes} corresponds to the distribution of agents in the two shelters, using parameter sets from the best-performing optimised individuals in 100 runs. In panels A and C, these results are obtained in setups with $50$ cockroaches, and no robots. In Panels B and D, the results are obtained in setups with $45$ cockroaches and $5$ robots.
Only results of the bifurcation diagram at specific values of $\sigma$ are shown. More generally, results before the bifurcation point ($\sigma < 0.8$) are similar to results at $\sigma = 0.4$, and results after the bifurcation point ($\sigma \geq 0.8$) are similar to results at $\sigma = 1.2$. 

Both the FSM and Hybrid models can be optimised to approximate correctly the decision-making dynamics corresponding to the ODE model  as shown in Fig.~\ref{fig:calibrationRes}.
Our methodology allows to generate many different parameter sets of the FSM and Hybrid models. Optimised parameter sets that correspond to the collective dynamics described by the ODE models can exhibit very different agent behaviour. In the case of the FSM model, the parameter $d$ of the diameter of the central zone of the arena is optimised: when this parameter is very low, the resulting agents ignore the wall-following behaviour of the FSM model.
In the FSM and in the Hybrid models, the parameters that influence the stopping behaviour ($s_{c,i,n}$, $\tau_{c,i,n}$, $\theta_i$) have fewer variation than the other parameters, with only a few islands of relevant values in the explored ranges.

\end{multicols}
\begin{multicols}{2}

We show that simulations performed with 45 cockroaches and 5 robots exhibit the same dynamics as the simulations of groups with 50 cockroaches and no robots (Fig.~\ref{fig:calibrationRes}). In this case the robots are governed by the same behavioral models as the insects but do not have the same parameter sets as those used for describing the natural behaviour of the cockroaches.
The detailed microscopic behaviours of the robots, \textit{e.g.} trajectories and movement patterns, can be very different from the microscopic behaviours of the animals. Nevertheless, we show that our methodology can be used to optimise the parameters of robot behavioural models in biohybrid systems to biomimic correctly the decision-making dynamics of the animals.

\end{multicols}

\begin{multicols}{2}

%% file: discussion.tex
\section{Discussion}
Few works in the animal and robotics literature try to tackle the problem of transitioning from models of one level of abstraction (reference model) to other another level of abstraction (target model).
Moreover, these studies only considered the transition between microscopic to macroscopic models~\cite{martinoli2004modeling,degond2008continuum,degond2013hierarchy}. 
The transition methodology adopted by these studies is incremental, and relies on the creation of intermediary models, dealing with both macroscopic and microscopic information, and that would share some parameters with the reference and target models. 
In~\cite{martinoli2004modeling}, this methodology is applied for modeling the behaviour of a swarm of autonomous robots for a collaborative task, from a microscopic model to a macroscopic one. The resulting model outperforms human-calibrated macroscopic models.
In~\cite{degond2008continuum}, a time-continuous kinetic mean field version of the Couzin-Vicsek model is obtained from its discrete microscopic version.
In~\cite{degond2013hierarchy}, continuous macroscopic models of the behaviour of pedestrian are obtained from discrete microscopic Agent-Based Models.
Additionally, no work in the literature investigates how to automate and generalise the transition process between models of different levels of abstraction. The transition process can also become more challenging if the reference and target models have no (or few) common parameter(s), or if their formulation is too different from each other.

Our methodology could also be relevant for problems of mixed-societies of animals and robots~\cite{halloy2007social}.
A mixed society is defined as a group of robots and animals able to integrate and cooperate: each robot is influenced by the animals, but can, in turn, influence the behaviour of the animals and of other robots. Individuals, natural or artificial, are perceived as equivalent, and the collective decision process results from the interactions between natural and artificial agents~\cite{halloy2007social,Gribovskiy2010,halloy2013384}.
In recent works robots in mixed-society have already been used to study individual and collective animal behaviours: robots are mixed with cockroaches in~\cite{sempo2006integration,halloy2007social}, chicks in~\cite{Gribovskiy2010,gribovskiy2015automated}, honeybees in~\cite{landgraf2010biomimetic}, fruit flies in~\cite{zabala2012simple}, guppies in~\cite{landgraf2014blending} and zebrafish in~\cite{bonnet2012437,polverino2013zebrafish,butail2014influence,butail2014fish}.
In this kind of systems, different approaches to modeling (macroscopic vs. microscopic; analytical vs. simulation) are strongly linked and complementary: different models deliver data necessary for the robot design process, provide explicit and analytic description of the observed collective behaviour, give predictions used for the society modulation, or ease the development of robot controllers.
Macroscopic models can convincingly describe collective dynamics, but cannot be implemented directly into robotic controllers. Robot controllers are intricately microscopic, as they describe the behaviour of individual agents. One of difficulties in experiments involving mixed societies is to implement the dynamics described in a macroscopic model into robot controllers (microscopic models). In previous mixed-societies studies (including~\cite{halloy2007social}), this process is done empirically.

More generally, complex systems exhibit multi-level dynamics (hierarchical organisation), with global and local behavioural patterns. Recent studies investigate the \textit{micro-macro link}, the relationship between macroscopic and microscopic descriptions of multi-level behavioural dynamics~\cite{hamann2008framework,reina2015quantitative,martinoli2004modeling,yamins2005towards}.  This problem also applies to the design of swarm group robotic controllers~\cite{hamann2008framework,martinoli2004modeling,lerman2005review,vigelius2014multiscale,mermoud2012real}.

\section{Conclusion}
Here, we tackle the problem of navigating between models of different levels of abstraction, in the context of animal collective decision-making.
Animals of collective behaviour can be macroscopic (analytical description of the behaviour of the population) or microscopic (explicit description of the behaviour and states of individuals and their interaction with the environment). Both type of models are complementary. Our methodology allows to transition from one to the other: we automate the design of a microscopic target model from a reference macroscopic model from the literature. We apply this methodology to the cockroaches shelter-selection problem described in~\cite{ame2006collegial,halloy2007social}. The ODE macroscopic model used as reference is described in~\cite{halloy2007social}.

We consider two target models. The FSM model~\cite{cazenille2015multi} is a microscopic agent-based model inspired from the literature on individual cockroach behaviour~\cite{jeanson2005selfOrganized,garnier2009self}. We introduce the Hybrid model, an agent-based model that uses both macroscopic and microscopic information. Both the FSM and Hybrid models can be used to replay the behaviour of animals in simulation, or be implemented as robot controllers.
We generate automatically parameter sets of the FSM and Hybrid models for cockroaches agents, calibrated to describe the same collective behaviour, and site-selection dynamics as the ODE model.

Our methodology is a first step toward generating the controllers of robots in a mixed-society of animals and robots. Mixing animals and robots can be useful to study animal behaviour, or even to modulate their individual or collective behaviour.

A subsequent study would include an application of this methodology to more complex setups, with more than two shelters and more than two population.
Finally, our methodology could be extended by generating microscopic Finite State Machine models from scratch, without \textit{a-priori} structural knowledge (i.e. the type and number of states).
Our methodology gives promising results, and could possibly be applied to model, calibrate, and modulate the collective behaviour of other species (\textit{e.g.} fishes, bees, or others).

%% file: Supplementary.tex
\section*{Supplementary Information}

\subsection*{\textbf{Simulations}}
Results from the FSM and Hybrid models are obtained from simulations of $28800$ time-steps, of the setup described in~\cite{ame2006collegial,halloy2007social} (cf Fig.~\ref{fig:ArenaCRExpe}): a circular arena (diameter $1m$) containing two identical shelters (diameter $150mm$).
For all models, only populations of $50$ individuals are considered (similar results are observed with populations of $16$ and $100$).

\subsection*{\textbf{ODE model resolution}}

The ODE model (from~\cite{halloy2007social}) used in this article is defined as follow:
\begin{equation} \label{eq:animalsEq}
	\frac{d x_{i}}{d t} =
		x_e \mu_i \left(1 - \frac{x_i + \omega r_i}{S_i} \right) -
		x_i \frac{\theta_i}{1 + \rho \frac{x_i + \beta r_i}{S_i}^n }
\end{equation}
\begin{equation} \label{eq:robotsEq}
	\frac{d r_{i}}{d t} =
		r_e \mu_{ri} \left(1 - \frac{x_i + \omega r_i}{S_i} \right) -
		r_i \frac{\theta_{ri}}{1 + \rho_r \frac{\gamma x_i + \delta r_i}{S_i}^{n_r} }
\end{equation}

\begin{equation} \label{eq:totEq}
	C = x_e + x_1 + x_2, \quad
	R = r_e + r_1 + r_2, \quad
	M = R + C
\end{equation}

Results are obtained by resolution of Eq.~\ref{eq:animalsRobots} using the Gillespie method~\cite{gillespie1977exact}.
The use of the Gillespie method for resolution allows experimental fluctuations to be taken into account. The Gillespie algorithm generates a birth-and-death stochastic process, described by the following master equation:
\begin{equation} \label{eq:masterEq}
	\begin{split}
	\frac{d}{d t} P(x_1, x_2, r_1, r_2, t) =
	& + W_1(x_1)P(x_1 - 1, x_2, r_1, r_2, t) - W_1(x_1)P(x_1, x_2, r_1, r_2, t) \\
	& + W_2(x_1 + 1)P(x_1 + 1, x_2, r_1, r_2, t) - W_2(x_1)P(x_1, x_2, r_1, r_2, t)\\
	& + W_3(x_2)P(x_1, x_2 - 1, r_1, r_2, t) - W_3(x_2)P(x_1, x_2, r_1, r_2, t)\\
	& + W_4(x_2 + 1)P(x_1, x_2 + 1, r_1, r_2, t) - W_4(x_2)P(x_1, x_2, r_1, r_2, t)\\
	& + W_5(r_1)P(x_1, x_2, r_1 - 1, r_2, t) - W_5(r_1)P(x_1, x_2, r_1, r_2, t) \\
	& + W_6(r_1 + 1)P(x_1, x_2, r_1 + 1, r_2, t) - W_6(r_1)P(x_1, x_2, r_1, r_2, t)\\
	& + W_7(r_2)P(x_1, x_2, r_1, r_2 - 1, t) - W_7(r_2)P(x_1, x_2, r_1, r_2, t)\\
	& + W_8(r_2 + 1)P(x_1, x_2, r_1, r_2 + 1, t) - W_8(r_2)P(x_1, x_2, r_1, r_2, t)
	\end{split}
\end{equation}

\begin{align}
	W_1		&=	x_e \mu_1 \left(1 - \frac{x_1 + \omega r_1}{S_1} \right) \quad & \quad
	W_2		&=	x_1 \frac{\theta_1}{1 + \rho \frac{x_1 + \beta r_1}{S_1}^n} \\
	W_3		&=	x_e \mu_2 \left(1 - \frac{x_2 + \omega r_2}{S_2} \right) \quad & \quad
	W_4		&=	x_2 \frac{\theta_2}{1 + \rho \frac{x_2 + \beta r_2}{S_2}^n} \\
	W_5		&=	r_e \mu_{r1} \left(1 - \frac{x_1 + \omega r_1}{S_1} \right) \quad & \quad
	W_6		&=	r_1 \frac{\theta_{r1}}{1 + \rho_r \frac{\gamma x_1 + \delta r_1}{S_1}^{n_r}} \\
	W_7		&=	r_e \mu_{r2} \left(1 - \frac{x_2 + \omega r_2}{S_2} \right) \quad & \quad
	W_8		&=	r_2 \frac{\theta_{r2}}{1 + \rho_r \frac{\gamma x_2 + \delta r_2}{S_2}^{n_r}}
\end{align}

This equation gives the time evolution of probability $\frac{d}{d t} P(x_1, x_2, r_1, r_2, t)$ to find $x_e$ animals outside the shelters, $r_e$ robots outside the shelters, $x_1$ animals under the first shelter, $r_1$ robots under the first shelter, $x_2$ animals under the second shelter and $r_2$ robots under the second shelter. As this probability depends only on the previous state of the system, the process is Markovian.
We calibrated the Gillespie algorithm to take into account the number of agent, as described in Eq.~\ref{eq:totEq}.\\

Only populations of $50$ individuals are considered (similar results are observed with populations of $16$ and $100$).

Halloy \textit{et al.}~\cite{halloy2007social} only consider cases with two populations and two sites.
The model could be generalised to $P$ sites and $N$ populations:
\begin{equation} \label{eq:animalsRobots}
\begin{split}
	\frac{d x_{j,k}}{d t} =
	x_e \mu_{j,k} \left(1 - \frac{\nu \cdot x^{\intercal}_{k}}{S_k}\right) -
	x_{j,k} \frac{\theta_{j,k}}{1+\rho_{j} \left(\frac{\alpha_{j} \cdot x^{\intercal}_{k}}{S_{j}}\right)^{n_j}} \\
	\: \mbox{for} \: j = 1,...,n \:\:\:\:\: k = 1, ..., p
\end{split}
\end{equation}

\begin{equation} \label{eq:totEq}
	M = x_e + \sum^{N}_{j} \sum^{P}_{k} x_{j,k}
\end{equation}

\subsection*{\textbf{Calibration of models} \label{sec:supplementaryCalibration}}
The simulated cockroaches agents using the FSM or the Hybrid model can be calibrated to exhibit the same collective behaviour the ODE model. As the ODE model is parametrised using experimental data, it allows the FSM and Hybrid model to describe as accurately as possible the (macroscopic) site-selection dynamics of the cockroaches. The calibration process is described in Fig.~\ref{fig:optim-task0}. \\

As there is only few \textit{a-priori} information about the parameter space, and as the parameter space has a relatively large dimensionality, we use the state-of-the-art CMA-ES evolutionary optimisation method (\cite{auger2005restart}, population size is 20, maximal number of generations is 500). The list of optimised parameters is found in Table~\ref{fig:parametersModels}.
All experiments were performed using the Grid'5000 computer grid (see https://www.grid5000.fr). Depending on the model and parameters set tested, each experimental runs can be performed in 1 to 15 hours in a 8-cores computer. 
\\

The fitness, minimised by CMA-ES, corresponds to a comparison between an optimised bifurcation diagram with the reference diagram from the ODE model.
It is computed as follow:
\newcommand{\fitness}{\text{Fitness}}
\newcommand{\hellinger}{\text{Hellinger}}
\newcommand{\optimized}{\text{optimised}}
\newcommand{\reference}{\text{reference}}
\newcommand{\calibration}{\text{calibration}}
\begin{equation}
	\fitness_{\calibration}(x) = D_{\hellinger}(B_{\optimized} / N_u, B_{\reference} / N_u)
\end{equation}
where $x$ is the tested parameter set (genome), $N_u$ is the number of considered values of $\sigma$ in the bifurcation diagrams ($10$) and $B_{\optimized}$ and $B_{\reference}$ are one-dimensional histograms version of the bifurcation diagrams. The term $N_u$ is used for normalisation.
The Hellinger distance~\cite{deza2006dictionary} is defined by the equation:
\begin{equation}\label{eq:hellinger}
D_{\hellinger}(P, Q) = \sqrt{ 2 \sum_{i=1}^{d} (\sqrt{P_i} - \sqrt{Q_i}  )^2 }
\end{equation}
where $P$ and $Q$ are two histograms, and $P_i$,$Q_i$ their $i$-th bins.
The Hellinger distance is a divergence measure, similar to the Kullback-Leibler (KL) divergence. However, the Hellinger distance is symmetric and bounded, unlike the KL-divergence (and most other distance metrics).
As such, it is adapted when comparing two histograms~\cite{deza2006dictionary}.